\documentstyle[aps]{revtex}
\begin{document}
\title{THE THREAT TO LIFE FROM ETA CARINAE AND GAMMA-RAY BURSTS} 

\author{Arnon Dar$^1$ and A. De R\'ujula$^2$}
\address {1. Department of Physics and Space Research Institute,
   Technion, Haifa 32000, Israel\\   
   2. Theory Division, CERN, Geneva 23, Switzerland}

\maketitle

\baselineskip=11.6pt
\begin{abstract}

{\bf Eta Carinae, the most massive and luminous star known in our galaxy,
is rapidly boiling matter off its surface. At any time its core could
collapse into a black hole, which may result in a gamma-ray burst (GRB)
that can devastate life on Earth. Auspiciously, recent observations
indicate that the GRBs are narrowly beamed in cones along the rotational
axis of the progenitor star. In the case of Eta Carinae the GRBs will not
point to us, but will be ravaging to life on planets in our galaxy that
happen to lie within the two beaming cones. The mean rate of massive life
extinctions by jets from GRBs, per life-supporting planet in galaxies like
ours, is once in 100 million years, comparable to the rate of major
extinctions observed in the geological records of our planet.}
\end{abstract}
\baselineskip=14pt

Gamma ray bursts (GRBs) are short-duration flares of MeV $\gamma$-rays
from outer space that last between a few milliseconds and $\sim$1000 s and
occur at a rate of about 3 a day \cite{MeeganF}. They were discovered
serendipitously in 1967 by the Vela satellites launched by the US to
monitor the compliance with the Nuclear Test Ban Treaty, banning nuclear
explosions in and above the atmosphere. Their exact locations ---and
consequently their distance and total energy output--- were unknown for 30
years, although their isotropy, established by observations with the BATSE
instrument on board the Compton Gamma Ray Observatory satellite (CGRO)
strongly suggested \cite{Meegan} cosmological distances \cite{Usov}.
Combined with the observed short-time variability of GRBs, such distances
imply an enormous energy release from a small volume, if due to spherical
explosions. Alternatively, it was argued that if GRB progenitors are so
distant, they must be produced by narrow relativistic jets, from the birth
of neutron stars or of black holes \cite{Woosley},\cite{ShavivD}.

The atmosphere is opaque to high energy $\gamma$-rays and cosmic ray nuclei,
and protects life on Earth from their incoming constant flux.  Collisions
in the upper atmosphere, however, produce a flux of energetic muons that
reach sea level, about $10^{-2}$ muons s$^{-1}$ cm$^{-2}$. Life on Earth,
apparently, has adjusted to the radiation damage from this small
flux of atmospheric muons, each depositing through ionization, in
biological materials, about $\rm  2.4\, MeV\, g^{-1}$. But,
if very large fluxes of $\gamma$-rays and cosmic ray nuclei suddenly impinge
on the atmosphere, they can have a devastating effect on life on Earth.  
In fact, it has been argued \cite{DLS} that the highly beamed 
cosmic rays from GRBs in our galaxy, that happen to point in our
direction, can produce lethal fluxes of atmospheric
muons at ground level, underground and underwater, destroy the ozone layer
and radioactivate the environment, so that GRBs could have caused some of
the massive life extinctions on planet Earth in the past 500 My.

Before 1997 the above arguments were mere speculation. But supporting
observational evidence accumulated after the significant
discoveries of long-lasting GRB X-ray, optical and radio ``afterglows'',
 made possible by the precise and prompt localization of GRBs by the
Italian--Dutch satellite BeppoSAX \cite{Costa}. The GRB and afterglow
observations have shown beyond doubt that the ``long'' duration GRBs
(which are the majority) take place in distant galaxies \cite{Metzger},
mainly in star formation regions, and are associated with supernova
explosions \cite{Galama}, \cite{Dado}.  Their large inferred energies, the
properties of their afterglows, their apparent association with
supernovae, and their global rate of $ \sim 1000$ per year, imply that
GRBs are highly beamed \cite{Dado},\cite{Dar}.

Because of the limited sizes of the satellite-borne detectors, GRBs have
been observed mostly in the sub-MeV energy region, where the photon number
flux, decreasing with increasing energy, is large enough. However, for a
few very bright GRBs the EGRET instrument on board the CGRO detected
$\gamma$-rays of up to GeV energies \cite{Hurley}. Moreover, four large
ground-based $\gamma$-ray detectors, the Tibet air shower array, the
HEGRA-AIROBICC \v Cerenkov array, the Milagro water-\v Cerenkov detector,
and GRANDE, have reported possible detections of TeV $\gamma$-rays in
directional and temporal coincidence with some GRBs detected by BATSE. In
every case, the estimated total energy in TeV photons was about 2 orders
of magnitude larger than the energy in sub-MeV photons measured by 
BATSE.
In particular, GRANDE \cite{Lin} and MILAGRITO \cite{Atkins} have reported
the detection of unexpectedly large fluxes of muons coincident in time and
direction with GRBs. These muons are allegedly produced by the
interactions in the upper atmosphere of $\gamma$-rays from the GRB with
energies well above 100 GeV.  These observations, if confirmed, would
imply that GRBs are more lethal than they were previously thought to be.
Since TeV photons are absorbed in the intergalactic infrared (IR)
background by pair production, only relatively close-by GRBs (for which
this absorption is insignificant) can be observed at TeV energies. This
may explain why only a small fraction of the BATSE-detected GRBs in the
fields of view of the various ground-based detectors were claimed to have
been seen at TeV energies.

In Table I we list the measured redshift and fluence $\rm F_\gamma$ (in
units of $10^{-5}$ erg cm$^{-2}$) in the BATSE energy band, 40--2000 keV,
for all GRBs with known redshift $\rm z$. We also list their inferred
luminosity distance $\rm D_L$ (in units of Gpc) and their total energy
output, $\rm E_\gamma = 4\, \pi\, D_L^2\, F_\gamma/(1+z)$  (in units of
$10^{53}$ erg), assuming isotropic emission and a critical Universe with a
Hubble constant $\rm H_0 = 65\, km\, s^{-1}\, Mpc^{-1}$, fractional matter
density $\rm \Omega_M=0.3$ and vacuum energy density $\rm
\Omega_\Lambda=0.7$.

Eta Carinae ---a large blue variable star in the Carina constellation,
more than 100 times as massive and 5 million times as luminous as the
Sun-- is one of the most massive and luminous stars known \cite{Davidson}.
It is rapidly boiling matter off its surface. At any time its core could
collapse into a black hole, which may result in a gamma-ray burst (GRB)
\cite{Woosley},\cite{ShavivD},\cite{Dado} . Should the violent end of Eta 
Carinae, the most massive star known in our galaxy and only $\rm D=2\, 
kpc$ away, emit
in our direction a GRB similar to that of the most energetic GRB in Table
I (GRB 990123), the atmosphere of Earth facing the star would be subject
to a total energy deposition:
\begin{equation} 
\rm{E_\gamma \over
4\,\pi\,D_L^2}\approx 4\times 10^{9}\,erg\,cm^{-2} 
\label{EG}
\end{equation} 
within seconds. This energy release is akin to that of the simultaneous
explosions in the upper atmosphere of one-kiloton of TNT per $\rm km^2$,
over the whole hemisphere facing Eta Carinae.  This would destroy the
ozone layer, create enormous shocks going down in the atmosphere, lit up
huge fires and provoke giant global storms.

If the energy 
of GRBs in TeV $\gamma$-rays, as indicated by various experiments
\cite{Lin},\cite{Atkins},  is $\sim 100$ times larger than
in the sub-MeV domain, the energy deposition of Eq. 1
would be correspondingly larger.
Moreover, the interactions of the TeV $\gamma$-rays in the upper atmosphere
would produce a lethal dose of highly penetrating muons,
destroying life on the surface, underground and underwater. Indeed, a
high energy $\gamma$-ray  impinging on the atmosphere
at a large zenith angle $\theta$ produces $\rm \sim
0.23\,\cos\theta\,[\epsilon_\gamma]^{1.17}$ muons at ground level 
\cite{Lin}, where $\rm
\epsilon_\gamma$ is the $\gamma$-ray energy in TeV. Hence, the total
muon fluence at ground level expected from a GRB from the supernova death
of Eta Carinae is  $\rm \sim 5\times 10^{10}\, cm^{-2}$ (the roughly
linear dependence on the $\gamma$-ray energy makes this result
sensitive only to the total deposited energy). The energy deposition 
by these high-energy muons in biological materials is 
$\rm \sim 2.5\times 10^5\, erg\,g^{-1}$, which is about ten times the 
lethal dose for human beings: the whole-body dose from
penetrating ionizing radiation resulting in 50\% mortality in 30 days
\cite{Groom}. 
 
All of the above, which would be devastating for life on Earth, 
would only happen if the
$\gamma$-rays from Eta Carinae's supernova point in our direction.
But would this GRB  point to us?  
There are at least three known superheavy stars in our galaxy
with a lifetime shorter than $\sim$ 1 My and expected to end in a giant
supernova, implying that the galactic rate of 
giant supernovae is $\rm \geq 3\times
10^{-6}\, y^{-1}$. The rate of massive life extinctions is $\rm \sim
10^{-8}\,~y^{-1}$. Thus, if all galactic giant supernovae produced deadly
GRBs, their $\gamma$-rays must be funnelled in a cone of opening angle $\rm
\theta_b\leq 5^o$, for two opposite GRBs per giant supernova. 
The chance probability for such cones to point in our
direction is only $3\times 10^{-3}$. But the expected direction for
a jetted GRB is the progenitor's polar axis, which for Eta Carinae 
points $\rm 57^o\pm 10^o$ away from our direction,  judging from the radial
velocities, proper motions and  projected shape of its equatorial disk 
of debris \cite{Davidson}. This reduces considerably the chance that the 
GRBs from
Eta Carinae point to our planet. Moreover, the properties of GRB afterglows
and their association with Type Ib/Ic supernovae
imply that GRBs  are beamed into much
narrower cones, of 1 mrad typical opening angle! \cite{Dado}. This
reduces to a negligible level the threat to terrestrial life from Eta Carinae.

Could Galactic GRBs beamed in our direction have caused 
some of the massive life extinctions in the history of Earth?
The average energy output of a GRB is 5 times smaller than 
that of GRB 990123, as can be seen in Table I. The average distance of 
galactic GRBs from Earth,
assuming they have the same spatial distribution as supernova remnants,
is $\sim 8$ kpc. Gamma rays alone from such ``typical'' GRBs can barely 
cause  major mass extinctions, since the
frequency of such GRBs is too small to explain a
mean rate of mass extinctions of once in $\sim$100 
My, observed in the geological records \cite{Benton}. 
However, if GRBs are produced in supernova explosions by highly 
relativistic jets of ``cannonballs'', as suggested by the striking success of the
Cannonball Model of GRBs in explaining their afterglows \cite{Dado},
the jetted cannonballs also produce highly beamed cosmic rays 
(CRs) by ionizing, sweeping up and accelerating the 
particles of the interstellar medium. 
Such CRs from galactic GRBs are much more devastating than their 
$\gamma$-rays. Let $\rm v$  be the speed of the CB and
$\rm\Gamma\equiv 1/\sqrt{1-(v/c)^2}\gg 1$ be its  Lorentz
factor. The bulk of the swept up ISM particles entering the CB with 
energy $\rm\Gamma\, m\, c^2$ in its rest frame are deflected by the CB's 
tangled magnetic fields, and are emitted isotropically in that frame.  In
the galactic rest frame their energy is Lorentz-boosted
to an average energy $\rm m\,c^2\,\Gamma^2$ 
and they are beamed into a cone of opening angle $\theta\sim
1/\Gamma$. Their energy distribution is related to the CBs' deceleration by
energy-momentum conservation, which yields $\rm dN_{CR}/d\Gamma\approx
N_{CB}/\Gamma^2$, where $\rm N_{CB}$ is the baryonic number of the CBs 
\cite{Dado}. The afterglows of the GRBs listed in Table I are very well 
fitted with 
initial Lorentz factors $\rm\Gamma_i\simeq 10^3$ and total baryonic number 
$\rm N_{jet}\sim  6\times 10^{50}$, comparable to that of the Earth
\cite{Dado}.  Thus, 
the energy fluence of CRs within their beaming cone of opening angle
$\rm\theta\leq\Gamma_i$, from a galactic GRB at a distance d $\sim 8$ kpc, is:
\begin{equation}
\rm F\simeq {E_{jet}\,\Gamma_i^2\over 3\,\pi\, d^2}
\simeq 1.5\times 10^{12}\, erg\, cm^{-2}\, .
\label{FGRB} 
\end{equation}
Most of this fluence is spread over less than $\rm \Delta t\sim $ 2 days, the 
typical CB deceleration time \cite{Dado} from $\rm\Gamma=\Gamma_i$ to
$\rm \Gamma=\Gamma_i/2$. It is carried by CRs with energies between
$\rm E= 2\, m_p\, c^2\Gamma_i^2\sim 2\times 10^3\, TeV$ and    
$\rm E=0.4\, m_p\, c^2\Gamma_i^2/4\sim 4\times 10^2\, TeV$\footnote{
The time delay of $10^3$ TeV protons relative to photons 
over ballistic trajectories of 8 kpc 
is only $\rm 8\, kpc/2\, c\, \gamma^2\simeq 0.41\, s.$}.

The ambient interstellar gas is transparent to the CR beam because the
Coulomb and hadronic cross sections are rather small with respect to typical
galactic column densities. Although the galactic magnetic field,
$\rm B\sim 5\times 10^{-6}\,Gauss$, results in a Larmor radius $\rm
r_{L}=\beta\,E_p /c\, q\,B\leq 10^{18}~cm\ll 8$ kpc for single protons 
with $\rm E_p\leq 10^{15}$ eV, it does not deflect and disperse the CR
beams from galactic GRBs.  This is because of the high collimation of the
CR beam which, even after travelling for a typical galactic distance
---e.g. d $\sim 8$ kpc, our distance from the Galaxy's centre---
has a very large energy  and pressure within an angle 
$\rm\theta \leq 1/\Gamma_i$ from its direction of motion:
$\rm E_{CR}\sim E_{jet}/3\sim 3\times 10^{51}\, erg$
 and $\rm P_{CR}\sim E_{jet}/(3\, \pi\, d^2\, c\, \Delta t ) \sim 3\times
10^{-4}\, erg\, cm^{-3}$, respectively. These figures are much larger than
the total magnetic energy of the swept-up galactic magnetic field inside
the cone, $\rm d^3\, B^2/24\, \Gamma_i^2 \sim 1.5\times 10^{49}$ erg and
the galactic magnetic pressure $\rm B^2/8\, \pi\sim 10^{-12}\,erg\,
cm^{-3}$.  Thus, the CR beam sweeps away the magnetic field along its way
and follows a straight ballistic trajectory through the interstellar
medium.  (The corresponding argument, when applied to the distant
cosmological GRBs, lead to the opposite conclusion: no CRs from 
distant GRBs accompany the arrival of gamma rays.)

The beam of multi-TeV cosmic rays accompanying a galactic GRB is deadly
for life on Earth-like planets. The total number of high energy muons
($\rm E_\mu\geq 25$ GeV)  in the atmospheric showers produced by a cosmic
ray proton with energy $\rm E_p \sim 10^2$ to $10^3$ TeV is $\rm
N_\mu(E>25\, GeV)\sim 9.14\, [E_p/TeV]^{0.757}/cos\theta$ \cite{Drees},
yielding a muon fluence at ground level:
\begin{equation} 
\rm F_\mu(E>25\, GeV)\simeq
1.7\times 10^{12}\, cm^{-2}\, .
\label{FMU} 
\end{equation}

Thus, the energy deposition rate at ground level in
biological materials, due to exposure to atmospheric muons produced by an
average GRB near the centre of the Galaxy, is $\rm 4.2\times 10^{12}\,
MeV\, g^{-1}$. This is approximately 270 times the lethal dose for 
human beings. The lethal dosages for other vertebrates and insects can be 
a few times or as much as a factor 20 larger, respectively. 
 Hence, CRs from galactic GRBs can produce a lethal dose of atmospheric
muons for most animal species on Earth.  Because of the large range of
muons ($\rm \sim 4 \,[E_\mu/GeV]\, m$ in water), their flux is lethal,
even hundreds of metres underwater and underground, for CRs arriving from
well above the horizon. Thus, unlike other suggested extraterrestrial
extinction mechanisms, the CRs of galactic GRBs can also explain massive
extinctions deep underwater and underground.  Although half of the planet
is in the shade of the CR beam, its rotation exposes a larger fraction of
its surface to the CRs, whose arrival time is spread over $\sim$ 2 days.
Additional effects increase the lethality of the CRs over the whole
planet. They include:

\noindent
(a) Environmental pollution  by radioactive
nuclei, produced by spallation of atmospheric and surface nuclei by 
the secondary particles of the CR-induced  showers. 

\noindent
(b) Depletion of stratospheric ozone, which reacts with the nitric
oxide generated by the CR-produced electrons 
(massive destruction of stratospheric ozone has been observed during large
solar flares, which generate energetic protons). 

\noindent
(c) Extensive damage to the food chain by radioactive pollution and
massive extinction of vegetation  by ionizing
radiation (the lethal radiation dosages for trees and plants are slightly
higher than those for animals, but still less than the flux given by
Eq.~3  for all but the most resilient species).

Are the geological records of mass extinctions consistent with 
the effects induced by cosmic rays from GRBs?
Good quality geological records, which extend up to $\sim 500$ My ago,
indicate that the exponential diversification of marine and continental life
on Earth over that period was interrupted by many extinctions 
\cite{Benton}, with the major ones ---exterminating more than 50\% of the 
species on land and sea--- occurring on average every
100 My. The five greatest events were those of the final
Ordovician period (some 435 My ago), the late Devonian (357
My ago), the final Permian  (251 My ago), the late Triassic 
(198 My ago) and the final Cretaceous (65 My ago).  The observed rate of
GRBs is $\rm \sim 10^3\, y^{-1}$. The sky density of galaxies brighter
than magnitude 25 (the observed mean magnitude of the host galaxies of the
GRBs with known redshifts) in the Hubble telescope deep field is $\sim
2\times 10^5$ per square degree \cite{Casertano}. Thus, the rate of 
observed GRBs,
per galaxy with luminosity similar to that of the Milky Way, is 
$ \rm R\sim1.2\times 10^{-7}\, y^{-1}$. To translate this result into the
number of GRBs born in our own galaxy, pointing to us, and occurring
at (cosmologically) recent times, one must take into account that
the GRB rate is
proportional to the star formation rate, which increases with redshift  like
$\rm (1+z)^3$ \cite{Reichart}.
For GRBs with known redshift (see Table I) one finds
$\rm\langle1+z\rangle\sim 2.1$. In a flat Universe (like ours) the probability
of a GRB to point to us within a certain angle is independent of distance.
Therefore, the mean rate of GRBs pointing to us and taking
place in our galaxy is roughly $\rm R/(1+z)^3\sim 1.3\times 10^{-8}\, y^{-1}$,
or once every $\sim$ 70 My. If most of these GRBs take place not
much farther away than the distance to the galactic centre, their effect
is lethal, and their rate is
consistent with the rate of the major mass extinctions on our planet in the
past 500 My.

The geological records also indicate that two of the major mass
extinctions were correlated in time with impacts of large meteorites or
comets, with gigantic volcanic eruptions, with huge sea regressions and
with drastic changes in global climate. A large meteoritic impact was
invoked \cite{Alvarez} in order to explain the iridium anomaly and the
mass extinction that killed the dinosaurs and claimed 47\% of existing
genera at the Cretaceous-Tertiary (K/T) boundary, 65 My ago. Indeed, a 180
km wide crater was later discovered, buried under 1 km of Cenozoic
sediments, dated back 65 My ago and apparently created by the impact of a
$\sim 10$ km diameter meteorite or comet near Chicxulub, in the Yucatan
\cite{Morgan}.  The huge Deccan basalt floods in India also occurred
around the K/T boundary 65 My ago \cite{Officer}.  The Permian/Triassic
(P/T) extinction, which killed between 80\% and 95\% of the species, is
the largest known in the history of life \cite{Erwin}; it occurred 251 My
ago, around the time of the gigantic Siberian basalt flood.  Recently,
possible evidence was found \cite{Becker} for a large cometary impact at 
that time.

The orbits of comets indicate that they reside in a spherical cloud at the
outer reaches of the solar system --the Oort Cloud \cite{Oort}-- with a
typical radius of $\rm R_O\sim 50000\, AU$. The statistics imply that it
may contain as many as $10^{12}$ comets with a total mass perhaps larger
than that of Jupiter. The large value of $\rm R_O$ implies that the comets
have very small binding energies and mean velocities of $\rm v\sim
100\,m\, s^{-1}$.  Small gravitational perturbations due to neighbouring
stars are believed to disturb their orbits, unbind some of them, and put
others into orbits that cross the inner solar system.  The passage of the
solar system through the spiral arms of the Galaxy where the density of
stars is higher, could also have caused such perturbations and
consequently the bombardment of Earth with a meteorite barrage of comets
over an extended period longer than the free fall time from the Oort cloud
to the Sun:
\begin{equation}
\rm t_{fall}=\pi \left [{R_O^3 \over 8\, G\, M_\odot}\right]^{1/2}
              \simeq 1.7\, My\, . 
\label{tfall} 
\end{equation} 
The impact of comets and meteorites from the Oort cloud could have
triggered the huge volcanic eruptions that created the observed basalt
floods, timed ---within 1 to 2 My--- around the K/T and P/T boundaries.
Global climatic changes and sea regression followed, presumably from the
injection of large quantities of light-blocking materials into the
atmosphere, from the cometary impacts and the volcanic eruptions. In both
the gigantic Deccan and Siberian basalt floods $\rm\sim 2\times 10^6~km^3$
of lava were ejected. This is orders of magnitude larger than in any other
known eruption, making it unlikely that the other major mass extinctions,
which are of a similar magnitude, were produced by volcanic eruptions. The
volcanic-quiet and impact-free extinctions could have been caused by GRBs.
Moreover, passage of the GRB jet through the Oort cloud after sweeping up
the interstellar matter on its way could also have generated
perturbations, sending some comets into a collision course with Earth,
perhaps explaining also the geologically active K/T and P/T extinctions.

The observation of planets orbiting nearby stars \cite{Mayor} has become
almost routine, but current techniques are insufficient to detect planets
with masses comparable to the Earth's. Future space-based observatories to
detect Earth-like planets are being planned. Terrestrial planets orbiting
in the habitable neighbourhood of stars, where planetary surface
conditions are compatible with the presence of liquid water, might have
global environments similar to ours, and harbour life. Our solar system is
billions of years younger than most of the stars in the Milky Way. Life on
extrasolar planets could have preceded life on Earth by billions of years,
allowing for civilizations much more advanced than ours. Thus Fermi's
famous question ``where are they?'', i.e. why did they not visit us or
send signals to us?  An answer is provided by GRB-induced mass
extinctions: even if advanced civilizations are not self-destructive, GRBs
can exterminate the most evolved species on any given planet or
interstellar vehicle at a mean rate of once every 100 My.  Consequently,
there may be no nearby aliens having evolved long enough to be capable of
communicating with us, or pay us a visit.

\acknowledgements{ 
The authors thank LeV Okun for useful comments. The partial support
by the Helen Asher Fund for Space Research and by the Technion VPR
Fund - Steiner Fund for the promotion of research is gratefully
acknowledged}.

\vskip 1.0 true cm
{\bf
\noindent
\centerline{
Table I - GRBs of known  redshift}}
\begin{table}[V]
\begin{tabular}{|l|c|c|c|c|}

GRB   &z & D$_{\rm L}$  & ${\rm F_\gamma}$
&${\rm E_\gamma}$\\
\hline
970228   &0.695  &4.55  &1.1   & 0.22  \\
970508   &0.835  &5.70  &0.32  & 0.07  \\
970828   &0.957  &6.74  &9.6   & 2.06  \\
971214   &3.418  &32.0  &0.94  & 2.11  \\
980425   &.0085  &.039  &0.44  &8.1E-6 \\
980613   &1.096  &7.98  &0.17  & 0.61  \\
980703   &0.966  &6.82  &2.26  & 1.05  \\
990123   &1.600  &12.7  &26.8  &19.80  \\
990510   &1.619  &12.9  &6.55  & 5.00  \\
990712   &0.434  &2.55  &6.5   & 0.53  \\
991208   &0.70   &4.64  &10.0  & 1.51  \\
991216   &1.020  &7.30  &19.4  & 5.35  \\
000131   &4.500  &44.4  & 4.2  &11.60  \\
000301c  &2.040  &17.2  &0.41  & 0.46  \\
000418   &1.119  &8.18  &2.0   & 0.82  \\
000911   &1.06   &7.66  &2.0   & 0.68  \\
000926   &2.066  &17.4  &2.20  &10.54  \\
010222   &1.474  &11.5  &12.0  & 7.80  \\
\end{tabular}
\end{table}
\vskip -0.3 true cm
\noindent
Redshift $\rm z$. Luminosity distance, $\rm D_L$, in Gpc.
Fluence measured by BATSE, $\rm F_\gamma$, in $10^{-5}$
erg cm$^{-2}$ units.
Deduced spherical energy,
$\rm E_\gamma$, in $10^{53}$ erg units.


\begin{thebibliography}{99}

\bibitem{MeeganF}
C.A.~Meegan, and G.J. Fishman, 
Ann. Rev. Astron. Astrophys. {\bf 33}, 415 (1995). 

\bibitem{Meegan}
C.A.~Meegan, {\it et al}, Nature  {\bf 355}, 143 (1992).

\bibitem{Usov}
V.V.~Usov, and G.V.~Chibisov, Astronomicheskii Zhurnal
 {\bf 52, No. 1},  192 (1975).

\bibitem{Woosley}
S.E.~Woosley,  Astrophys. Jour. {\bf 405}, 273 (1993).

\bibitem{ShavivD}
N.~Shaviv and A.~Dar, Astrophys. Jour. {\bf 447}, 863 (1995).

\bibitem{DLS}
A.~Dar, A.~Laor and N.~Shaviv, Phys. Rev. Lett. {\bf 80}, 5813 (1998).

\bibitem{Costa}
E.~Costa {\it  et al.}, Nature {\bf 387} 783 (1997).

\bibitem{Metzger}
M.R.~Metzger, {\it et al.}, Nature, {\bf 387} 878 (1997).

\bibitem{Galama}
T.J.~Galama, Nature  {\bf 395}, 670 (1998).

\bibitem{Dado}
S.~Dado, A,~Dar, and A.~De R\'ujula, astro-ph/0107367 (2001) 
and references therein.

\bibitem{Dar}
A.~Dar, Astrophys. Jour. {\bf 500} L93 (1998). 

\bibitem{Hurley}
K.~Hurley, Nature {\bf 372}, 652 (1994).

\bibitem{Lin}
T.F.~Lin {\it et al.}, {\bf ICRC26 Vol. 4}, 24 (1999).

\bibitem{Atkins}
R.~Atkins {\it et al.}, Astrophys. Jour. {\bf 533}, L119 (2000).

\bibitem{Davidson}
K.~Davidson, and R.M.~Humphrey, Ann. Rev. Astron. Astrophys. {\bf 
35}, 1 (1997). 

\bibitem{Groom}
D.E.~Groom {\it et al.}, Europ. Phys. Jour. {\bf C15}, 1 (2000)

\bibitem{Benton}
M.J.~Benton, Science, {\bf 268}, 52 (1995).

\bibitem{Drees}
M.~Drees, F. Halzen and K. Hikasa, Phys. Rev. {\bf D39}, 1310 (1989).

\bibitem{Casertano}
S.~Casertano  {\it et al.},  Astron. Jour. {\bf 120}, 2747 (2000).

\bibitem{Reichart}
D.E.~Reichart {\it et al.}, Astrophs. Jour. {\bf 552}, 57 (2001). 

\bibitem{Alvarez}
L.W.~Alvarez {\it et al.}, Science {\bf 208}, 1095 (1980).

\bibitem{Morgan}
A.R.~Hildebrand, and G.T. Mexico, Eos, {\bf 71},1425 (1990);
J.~Morgan {\it et+ al},  Nature  {\bf 390} 472 (1997).

\bibitem{Officer}
C.B.~Officer {\it et al.}, Nature {\bf 326}, 143 (1987);
V. Courtillot, {\it et al.}, Nature, {\bf 333}, 843 (1988);
V. Courtillot, Scientific American, {\bf 263}, October, 53 (1990);
C. B. Officer \& J. Page, {\it The Great Dinosaurs Controversy}
Addison Wesley (1996).

\bibitem{Erwin}
D.H.~Erwin, Nature, {\bf 367}, 231 (1994);
D.H.~Erwin, Scientific American {\bf 275}, July 56  (1996).

\bibitem{Becker}
L.~Becker {\it et al.}, Science {\bf 291}, 1530 (2001).

\bibitem{Oort}
J.H.~Oort, Bull. Astr. Inst. Neth. {\bf 11}, 91 (1950).

\bibitem{Mayor}
M.~Mayor and D.~Queloz, Nature, {\bf 378}, 355
(1995); R.P.~Butler and G.W.~Marcy, Astrophys. Jour. {\bf 464}, L153 
(1996).



\end{thebibliography}
\end{document}